\documentclass[sigconf]{acmart}

\copyrightyear{2024}
\acmYear{2024}
\setcopyright{acmlicensed}
\acmConference[CIKM '24]{Proceedings of the 33rd ACM International Conference on Information and Knowledge Management}{October 21--25, 2024}{Boise, ID, USA} 
\acmBooktitle{Proceedings of the 33rd ACM International Conference on Information and Knowledge Management (CIKM '24), October 21--25, 2024, Boise, ID, USA}
\acmDOI{10.1145/3627673.3680095} 
\acmISBN{979-8-4007-0436-9/24/10}

\AtBeginDocument{%
  }

\usepackage{url}            
\usepackage{nicefrac}       
\usepackage{microtype}      
\usepackage{algorithm}  
\usepackage{algpseudocode}
\usepackage{graphicx}
\usepackage{enumitem}
\usepackage{multirow}
\usepackage{xcolor}

\begin{document}

\title{Relevance Filtering for Embedding-based Retrieval}



\author{Nicholas Rossi}
\affiliation{%
  \institution{Walmart Global Technology}
  \city{Sunnyvale}
  \country{USA}
}
\email{nicholas.rossi@walmart.com}

\author{Juexin Lin}
\affiliation{%
  \institution{Walmart Global Technology}
  \city{Sunnyvale}
  \country{USA}
}
\email{juexin.lin@walmart.com}

\author{Feng Liu}
\affiliation{%
  \institution{Walmart Global Technology}
  \city{Sunnyvale}
  \country{USA}
}
\email{f.liu@walmart.com}

\author{Zhen Yang}
\affiliation{%
  \institution{Walmart Global Technology}
  \city{Sunnyvale}
  \country{USA}
}
\email{zhen.yang@walmart.com}

\author{Tony Lee}
\affiliation{%
  \institution{Walmart Global Technology}
  \city{Sunnyvale}
  \country{USA}
}
\email{tony.lee@walmart.com}

\author{Alessandro Magnani}
\affiliation{%
  \institution{Walmart Global Technology}
  \city{Sunnyvale}
  \country{USA}
}
\email{alessandro.magnani@walmart.com}

\author{Ciya Liao}
\affiliation{%
  \institution{Walmart Global Technology}
  \city{Sunnyvale}
  \country{USA}
}
\email{ciya.liao@walmart.com}
\renewcommand{\shortauthors}{Rossi et al.}

\begin{abstract}
In embedding-based retrieval, Approximate Nearest Neighbor (ANN) search enables efficient retrieval of similar items from large-scale datasets. While maximizing recall of relevant items is usually the goal of retrieval systems, a low precision may lead to a poor search experience. Unlike lexical retrieval, which inherently limits the size of the retrieved set through keyword matching, dense retrieval via ANN search has no natural cutoff. Moreover, the cosine similarity scores of embedding vectors are often optimized via contrastive or ranking losses, which make them difficult to interpret. Consequently, relying on top-K or cosine-similarity cutoff is often insufficient to filter out irrelevant results effectively. This issue is prominent in product search, where the number of relevant products is often small. This paper introduces a novel relevance filtering component (called "Cosine Adapter") for embedding-based retrieval to address this challenge. Our approach maps raw cosine similarity scores to interpretable scores using a query-dependent mapping function. We then apply a global threshold on the mapped scores to filter out irrelevant results. We are able to significantly increase the precision of the retrieved set, at the expense of a small loss of recall. The effectiveness of our approach is demonstrated through experiments on both public MS MARCO dataset and internal Walmart product search data. Furthermore, online A/B testing on the Walmart site validates the practical value of our approach in real-world e-commerce settings.
\end{abstract}



\maketitle

\section{Introduction}

Dense retrieval models \cite{bromley1993signature, karpukhin2020dense}, have greatly improved information retrieval by learning to represent queries and documents as dense vectors, capturing semantic relationships. This enables embedding-based retrieval to efficiently retrieve semantically relevant documents using Approximate Nearest Neighbor (ANN) search \cite{JDH17}. 
This approach has proven successful across a wide range of domains, including web search \cite{huang2013learning, shen2014latent}, question answering \cite{severyn2015learning}, and e-commerce search \cite{lin2024enhancing, magnani2022semantic, wang2021personalized, zhang2020towards, he2023que2engage}. While dense retrieval models are effective at retrieving relevant documents, their emphasis on recall can compromise precision by surfacing irrelevant documents. This issue is prominent in e-commerce product retrieval, where the number of relevant products is often small. Presenting a long list of irrelevant products may frustrate users, even if the relevant products are on top.

Traditionally, retrieval and reranking are treated as separate tasks, and the latter focuses on optimizing precision. However, as the number of retrieved documents increases, the computational burden on the reranker to demote irrelevant products may grow unnecessarily. For instance, if there is only one relevant product, retrieving $K$ documents from dense retrieval will include at least $K-1$ irrelevant ones. Hence, we propose to filter out the irrelevant products before reranking to enhance precision with negligible sacrifice of recall.


In principle, one could filter out the documents whose cosine similarity score is below a global threshold. However, the cosine similarity scores are usually not interpretable and should not be compared across different queries. Thus, applying a global threshold to the cosine similarity score is not an optimal approach.

To address this, we propose a "Cosine Adapter", which is a component in the neural network that provides a function to transform the cosine similarity score into an interpretable relevance score, which can be compared across queries. It is now valid to apply a global threshold to the relevance score to filter out low-relevance documents. Importantly, the Cosine Adapter's output function is query-dependent, which enables contextual awareness of query difficulty.

Our relevance mapping technique offers two key advantages: First, it provides a clear probabilistic interpretation of document scores, enhancing transparency and interpretability. Second, it is computationally efficient, as the mapping is applied directly to the cosine similarity scores. Extensive offline benchmarking on both MS MARCO and Walmart product search datasets, along with live online tests, demonstrate improved precision across diverse queries and applications.

\section{Related Work}
There has been previous work on \emph{ranked list truncation}, which is the problem of determining where to truncate a list of retrieved documents \cite{arampatzis2009,lien2019,bahri2020,wang2022,ma2022,zamani2022}. The idea is that the sequence of document scores (e.g., BM25 scores) and document statistics may have learnable patterns in terms of where the appropriate cutoff position is. For example, a large drop in document score may signal that the remaining documents are less relevant and the optimal cutoff is here. A model is trained to predict the cutoff position to optimize an evaluation metric (e.g., F1) of the truncated list. Recent works include a self-attention layer in the model to capture long-range dependencies among documents \cite{bahri2020,wang2022,ma2022}. Our approach is quite different: we utilize the query embedding, which contains a lot of information about the query, to learn what the cosine score of a relevant item may be. Our approach is also more computationally efficient, as discussed in Section~\ref{section.workflow}.

Embedding-based retrieval has recently been widely adopted in production systems across various companies \cite{nigam2019semantic, yang2020mixed, zhang2020towards, huang2020embedding, yao2021learning}. While numerous successes have been reported, the challenge of controlling relevance for embedding-based retrieval system remains an active area of research \cite{lin2024enhancing, li2021embedding}. Furthermore, there is growing interest in optimizing the pre-ranking phase for enhanced efficiency and scalability. This involves using lightweight modules to filter out less promising candidates before the more computationally intensive ranking stage \cite{wang2020cold, xu2024optimizing}. 
However, these works do not primarily focus on improving retrieval relevance. To enhance the relevance control in the pre-ranking phase, one approach leverages lexical matching, where a relevance control module filters out products based on the presence of key query terms in their titles \cite{li2021embedding}. A drawback of this approach is that it sacrifices some of the advantages of dense retrieval, which does not rely on text match. Research addressing both the efficiency and effectiveness of relevance control during pre-ranking remains limited.

\section{Background}
\subsection{Embedding-based retrieval}
In information retrieval, the objective is to identify the relevant candidates within a corpus for a given query. Candidates can be passages or documents. In embedding-based retrieval, the dual encoder architecture is used for this task, where two encoders convert the query and candidate, respectively, into a $d$-dimensional embedding vector \cite{karpukhin2020}. The relevance of a candidate to the query is measured by the cosine similarity between their embeddings. Once trained, the encoder is used to generate embeddings for all candidates in the corpus, forming an index. When the user searches for a query, the query embedding is generated, and an ANN search is conducted to retrieve the top-K candidates with the highest cosine similarity scores. 

A popular approach for training dual encoder models uses a contrastive loss and in-batch negative sampling \cite{chen2017sampling, henderson2017efficient, karpukhin2020, liu2021que2search, dong2022exploring}. Given a query $q_i$ and its corresponding positive candidate $p_i$, the contrastive learning loss is formulated as:
\begin{equation} \label{eq.contrastive_loss}
loss_i = - \log \frac{\exp{\left(\cos{\left(q_i, p_i \right)}/\tau\right)}}{\exp{\left(\cos{\left(q_i, p_i \right)}/\tau\right)} + \sum_{j\in\mathcal{N}} \exp{\left(\cos{\left(q_i,p_j\right)}/\tau \right)}},
\end{equation}
where $\tau$ is a temperature hyperparameter, and $\mathcal{N}$ denotes the set of negatives sampled within the batch.

An alternative approach is to use a softmax listwise loss \cite{lin2024enhancing, magnani2022semantic, zheng2022multi}. It leverages the predefined labels for each query-candidate pair. For a given query $q_i$ and a set of candidates $\mathcal{P}_i$, the ranking objective is defined as: 
\begin{equation} \label{eq.softmax_loss}
loss_i = - \sum_{j \in \mathcal{P}_i} y_{ij} \log \frac{\exp{\left(\cos{\left(q_i, p_j \right)}/\tau\right)}}{\sum_{j \in \mathcal{P}_i} \exp{\left(\cos{\left(q_i,p_j\right)}/\tau \right)}},
\end{equation}
where $y_{ij}$ is the predefined label, and $\tau$ is a temperature hyperparameter.

Contrastive loss primarily aims to distinguish the positive from the negatives, while softmax listwise loss aims to assign higher scores to candidates with higher labels. Despite their distinct formulations, both loss functions shape the embedding space based on the relative distances between candidates, rather than their absolute relevance scores. This makes it difficult to interpret the cosine similarity scores. Since the embedding space is optimized for relative comparisons within a query, cosine similarity scores cannot be compared across different queries.

\subsection{Problem statement}
\label{section.problem}

To minimize the impact on retrieval performance, we decouple relevance filtering from the retrieval process itself, performing filtering as a post-processing step on the ANN search results. Effective relevance filtering requires scores that represent relevance in an absolute sense. 

Given a query $q_i$ and its top-$K$ retrieved set $\mathcal{P}_i$ from ANN search, our goal is to identify a subset of candidates $\tilde{\mathcal{P}}_i$ that are truly relevant to the query. To achieve this, we introduce a function designed to measure the absolute relevance of each candidate and apply a cutoff threshold for filtering. 

A straightforward approach would be to employ a classifier to predict the relevance scores of query-candidate pairs \cite{dai2019deeper}. However, this approach is often computational expensive to deploy online, particularly when the number of retrieved candidates ($K$) is large.

We propose a simpler function $\mathcal{F}$ that operates directly on the existing cosine similarity scores from the retrieval system. Since our approach leverages the output of the existing dual encoders, it is cheap to compute. The parameters of $\mathcal{F}$ are query-dependent, learned from a neural network that takes the query embedding as input, as discussed below. Formally, the filtering process can be expressed as
\begin{equation} 
\tilde{\mathcal{P}}_i =  \{p_j | \mathcal{F}_{\Theta} (\cos(q_i, p_j) ) \geq t\},
\end{equation}
where $\Theta$ represents the query-dependent parameters of $\mathcal{F}$, and $t$ is a global threshold determined empirically.  


\section{Methodology}
\subsection{Cosine Adapter}
\begin{figure}
\centering
\includegraphics[width=0.48\textwidth]{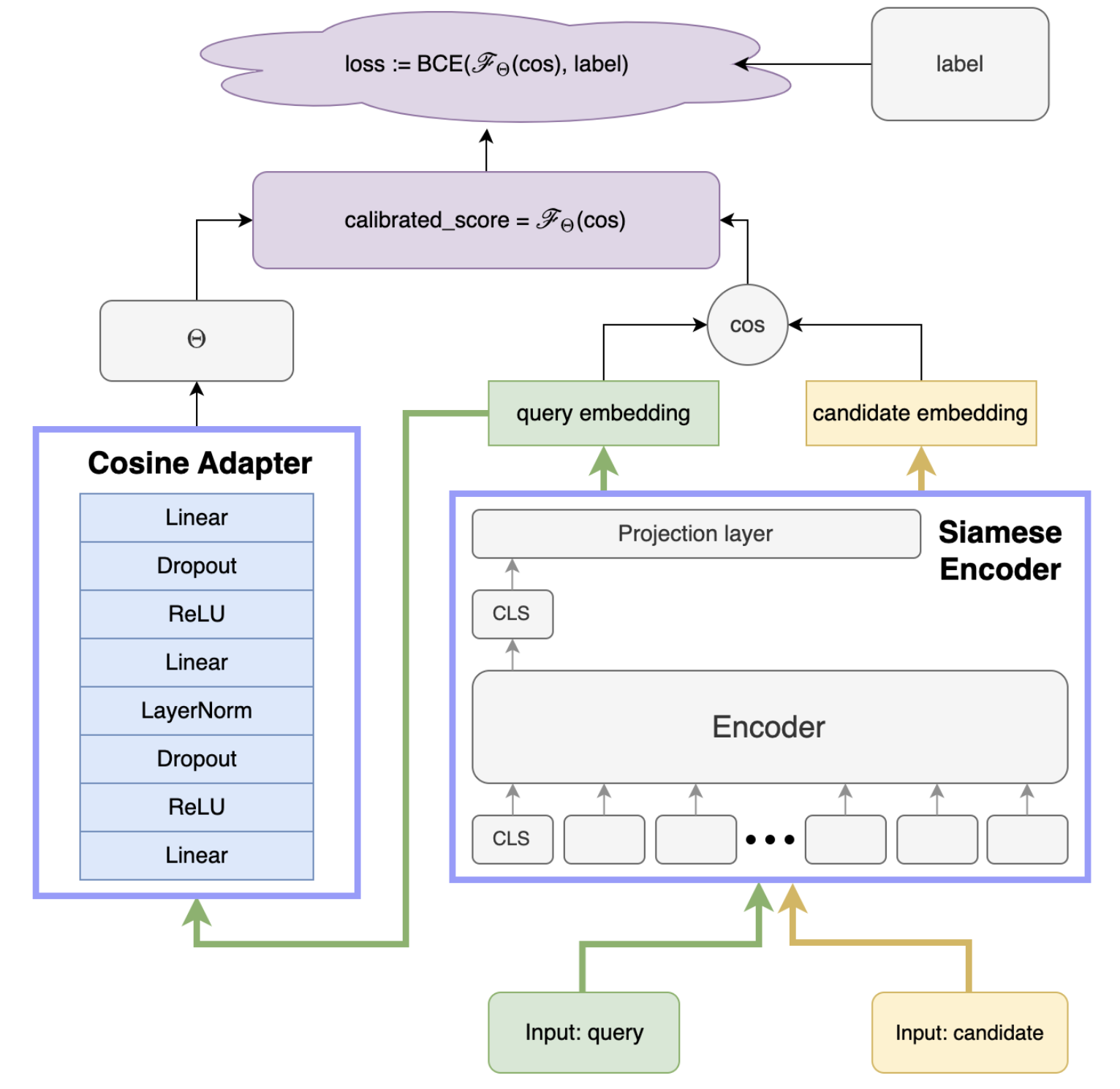}
\caption{Architecture of the Cosine Adapter. Siamese dual encoder is frozen during training.}
\label{fig.guardrail_model} 
\end{figure}

We explore several transformation functions designed to calibrate cosine similarity scores (which range from -1 to 1) to interpretable scores. To preserve the relative ranking of candidates and minimize the impact on recall performance, these functions are chosen to be monotonic and exhibit a variety of shapes:


\begin{enumerate}
\setlength\itemsep{0.5em}
    \item raw score: $\mathcal{F} (x) = x$
    \item linear: $\mathcal{F} (x | \Theta=(a,b)) = a x + b$
    \item square root: $\mathcal{F} (x | \Theta=(a,b))$ = $\text{sgn}(x) a \sqrt{|x|} + b $
    \item quadratic: $\mathcal{F} (x | \Theta=(a,b))$ = $\text{sgn}(x) a x^2 + b $
    \item power: $\mathcal{F} (x | \Theta=(a,b,k))$ = 
     $\text{sgn}(x) a |x|^k + b, $
     where $k \in (0,2)$.
\end{enumerate}
Function (1) is a baseline for comparison. The parameters $a$, $b$, and $k$ in these functions are query-dependent and learned by a neural network, which we call the "Cosine Adapter". As depicted in Figure~\ref{fig.guardrail_model}, the Cosine Adapter takes the query embedding from the dual encoder as input and outputs the parameter set $\Theta$. The Cosine Adapter involves a few feedforward layers with ReLU activation. For (5), $k$ is constrained to the range (0, 2) via a "$2 \sigma()$" transformation.

The raw cosine similarity score from a dual encoder model is transformed into a calibrated score using the corresponding $\mathcal{F}$ and $\Theta$. We train the Cosine Adapter layers via binary cross-entropy loss \cite{mackay2003information}:
\begin{equation} \label{eq.loss}
    loss_i = - \left[ y_i \log(\sigma(\mathcal{F}_i)) + (1-y_i) \log(1-\sigma(\mathcal{F}_i)) \right],
\end{equation}
where $y_i$ represents the label for the $i$-th query-candidate pair, and $\sigma$ denotes the sigmoid function. $\mathcal{F}_i$ and $\sigma(\mathcal{F}_i)$ are the logit and probability, respectively, that the query-candidate pair is relevant.

Note that in our proposal, we freeze the dual encoders while training the Cosine Adapter. Since the Cosine Adapter depends on the dual encoder and the relevance training data, it needs to be trained specifically for each dataset. Also, the training data for the Cosine Adapter can be different from that of the dual encoder.

\subsection{Relevance filtering workflow}
\label{section.workflow}
\begin{figure*}
\centering
\includegraphics[width=0.8\textwidth,height=0.35\textheight]{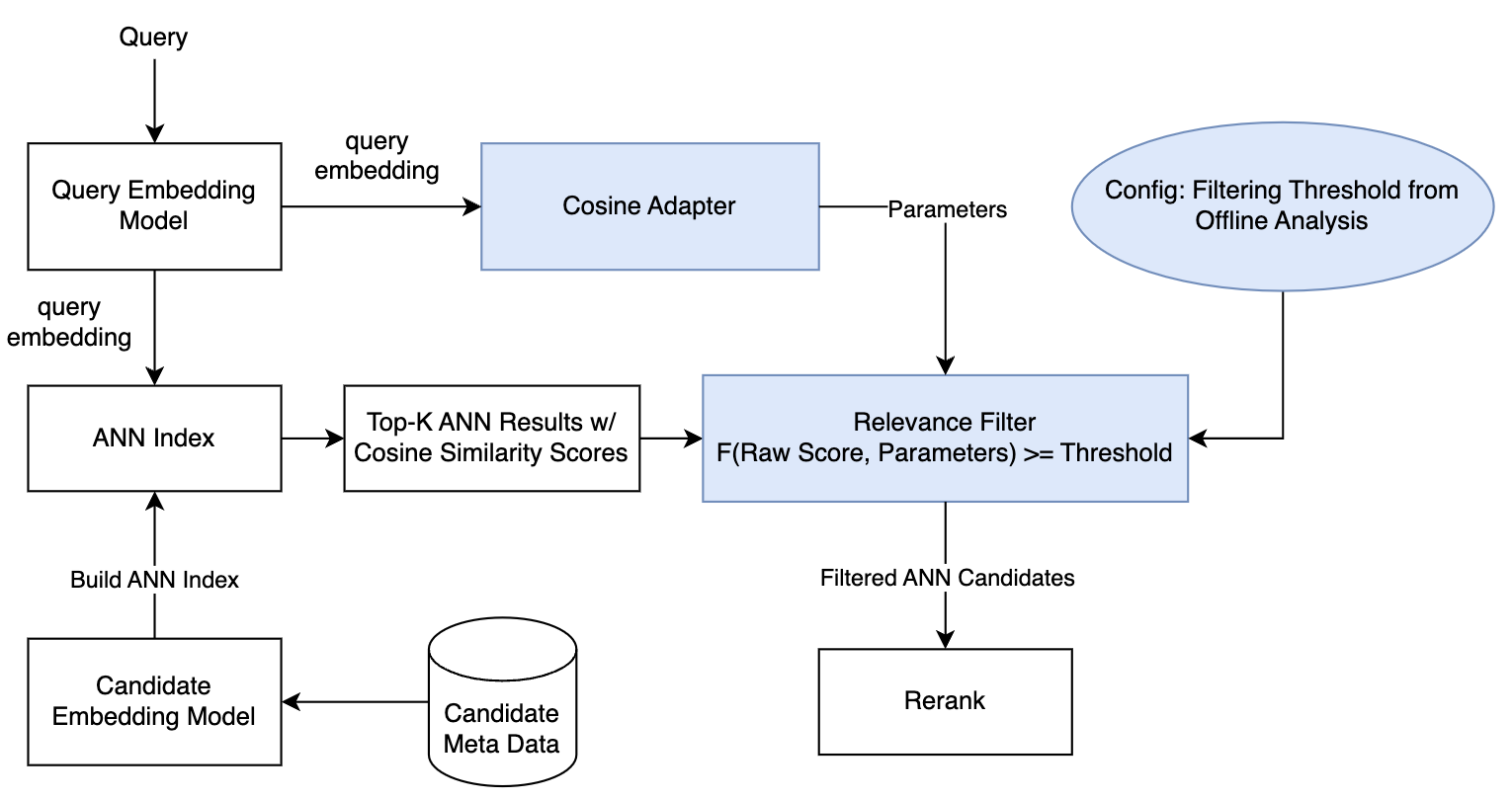}
\caption{Workflow of embedding-based retrieval with relevance filtering module. The components highlighted in blue represent the newly introduced elements for relevance filtering.}
\label{fig.workflow} 
\end{figure*}

Figure~\ref{fig.workflow} illustrates the integration of the Cosine Adapter and relevance filtering module into the retrieval system. Given a query, the query encoder generates the query embedding. The Cosine Adapter converts the query embedding into the parameters $\Theta$. The query embedding is sent to the ANN index to retrieve the top-$K$ candidates, along with their cosine similarity scores. Within the relevance filter component, the calibrated scores are calculated based on the transformation function and the raw cosine similarity scores. A global threshold, learned offline, is then applied to filter out the results. Finally, the filtered results are forwarded to the re-ranking module.

At inference time, the computational complexity involves the feedforward layers of the Cosine Adapter and computing the calibrated scores. The complexity of the feedforward layers is $O(d^2)$, where $d$ is the embedding dimension; note that the Cosine Adapter is run only once per query. The complexity of computing the calibrated scores is $O(K)$, since it involves a few operations per candidate. For comparison, some ranked list truncation approaches (like Choppy) \cite{bahri2020,wang2022,ma2022} include a self-attention layer among candidates, which has computational complexity of $O(K^2 d)$ \cite{vaswani2017}.

\section{Experiments and Results}
To demonstrate the effectiveness of our proposed method, we conduct experiments on both the public MS MARCO dataset and a private Walmart product search dataset. Furthermore, we integrate the relevance filtering component into Walmart's embedding-based retrieval system to validate its real-world utility. 
\subsection{Metrics and baselines}
Implementing a filtering mechanism in retrieval introduces a trade-off between recall and precision. We evaluate performance using the following metrics:
\begin{itemize}
    \item \textbf{PR AUC} (Area Under Precision-Recall Curve): This is a common metric for classification problems, which captures the overall balance between precision and recall across different thresholds. This metric is computed without filtering being implemented.
    \item \textbf{P@R95} (Precision at 95\% Recall): We establish a global cutoff threshold, applicable to all queries, that achieves 95\% recall relative to no filtering. We then report the corresponding precision. This metric highlights the precision achieved while maintaining a high level of recall.
    \item \textbf{Filter\%}: This is the percentage of results filtered out.
    \item \textbf{Null\%}: This is the percentage of queries that return zero results after filtering.
    \item \textbf{MRR} (Mean Reciprocal Rank). We report this for experiments on the MS MARCO dataset, since it is a standard metric. 
\end{itemize}
These metrics provide a comprehensive view of the filtering mechanism's impact on retrieval performance.

We report several baselines for comparison. We apply a global threshold on the raw cosine score ($\mathcal{F} (x) = x$). We also apply a global threshold to the max-normalized cosine score, i.e., the cosine score is normalized by the maximum score for the query. For comparison with recent work on ranked list truncation, we also include results for Choppy, optimizing for F1 \cite{bahri2020}.
\subsection{Experiment on MS MARCO data}

\begin{table*} 
\centering
\begin{tabular}
{c|l|c|c|c|c|c}
 \hline
 K & Method  & PR AUC & P@R95  & Filter\% & Null\%  & MRR\\ 
 \hline
  \hline
 \multirow{6}{*}{10}
  & no filter & - & P=0.0730, R=0.6973 & - & - & 0.411\\ 
  & Choppy & - & P=0.2775, R=0.2701 & 90.00 & 0.00 & 0.278\\ 
  & raw score & 0.124 & 0.0830 & 16.53 & 2.02 & 0.400\\ 
  & raw score (max-norm) & 0.194 & \textbf{0.0926} & \textbf{25.12} & 0.00 & 0.404 \\ 
 \cline{2-7}
 & linear & 0.206 & 0.0864 & 19.84 & 2.13 & 0.402\\ 
 & square root  & \textbf{0.208} & 0.0870 & 20.40 & 2.22 & 0.402 \\ 
 & quadratic & 0.205 & 0.0869 & 20.29 & 2.18 & 0.402\\ 
 & power  & 0.207 & 0.0871 & 20.36 & 2.15 & 0.402\\ 
 \hline
 \multirow{6}{*}{1000}
 & no filter & - & P=0.0011, R=0.9869 & - & - & 0.422\\ 
 & raw score & 0.041 & 0.0028 & 64.94 & 0.60 & 0.417\\ 
 & raw score (max-norm) & 0.119 & 0.0047 & 78.92 & 0.00 & 0.419\\ 
  \cline{2-7}
 & linear & 0.139 & 0.0065 & 84.63 & 0.09 & 0.421\\ 
 & square root & \textbf{0.141} & 0.0063 & 84.08 & 0.10 & 0.421 \\ 
 & quadratic & 0.138 & 0.0064 & 84.43 & 0.13 & 0.421\\ 
 & power & 0.139 & \textbf{0.0066} & \textbf{84.90} & 0.04 & 0.420\\ 
\hline
 \end{tabular}
\caption{Experiments on MS MARCO passage ranking dataset. The best scores are shown in boldface. "P" and "R" denote precision and recall, respectively.}
\label{table.msmarco}
\end{table*}

We conduct our experiment on the MS-MARCO passage ranking dataset \cite{nguyen2016ms}. This dataset, based on Bing search results, comprises 503k training queries and 8.8 million passages. As the test set labels are not publicly available, we train our model on the training set and report our results on the development set, which comprises 6,980 queries.
\subsubsection{Implementation}
We use the publicly available 
simlm-base-msmarco-finetuned \footnotemark{}, a fine-tuned SimLM model trained on the MS-MARCO dataset using contrastive loss and knowledge distillation \cite{wang2022simlm}, as the dual encoder in our experiments. We train the Cosine Adapter with a 1:31 positive-to-negative ratio, where negatives are mined from BM25 results provided by \cite{wang2022simlm}. The model is trained on 2 A100 GPUs with batch size 128 for 3 epochs. The dual encoder is frozen during this training.
\footnotetext{https://huggingface.co/intfloat/simlm-base-msmarco-finetuned}

\subsubsection{Results}

We evaluate the impact of our filtering mechanism on retrieval sets of varying sizes ($K = 10, 1000$), with results presented in Table~\ref{table.msmarco}. Notably, PR AUC and precision values are generally low, as most evaluated queries have only one relevant passage.

The our proposed methods outperform the baselines in terms of PR AUC, indicating a stronger ability to distinguish relevant passages. For $K=1000$, our methods consistently yield higher P@R95 and Filter\% values, indicating better filter performance. Among the proposed functions, the power function outperforms the others, although the differences are small.

Finally, our proposed methods significantly reduce the number of queries with zero results (Null\%) compared to the raw-score baseline when $K = 1000$. This indicates that our methods are more effective at identifying the correct boundary between relevant and irrelevant results, rather than indiscriminately removing all results. Interestingly, when $K = 10$, filtering on raw scores exhibits a lower Null\%. This is due to the fact that 30\% of the queries lack positive results within the top 10, whereas this percentage drops to 1\% when $K = 1000$. Further analysis reveals that among queries yielding zero results with the power method but at least one result with the raw-score baseline, about 70\% do not contain positive results within the top 10. This observation reinforces that our proposed methods are more adept at filtering irrelevant results.

We find that Choppy has a relatively low recall because it consistently truncates the list too aggressively. This may be due to the peculiar nature of the dataset, i.e., the fact that there is usually only 1 relevant passage per query. Note that PR AUC and P@R95 are not reported for Choppy, because it does not have a tunable threshold: a single cutoff position is predicted for each query \cite{bahri2020}.

\subsubsection{Analysis}
\begin{figure}
\centering
\includegraphics[width=0.45\textwidth]{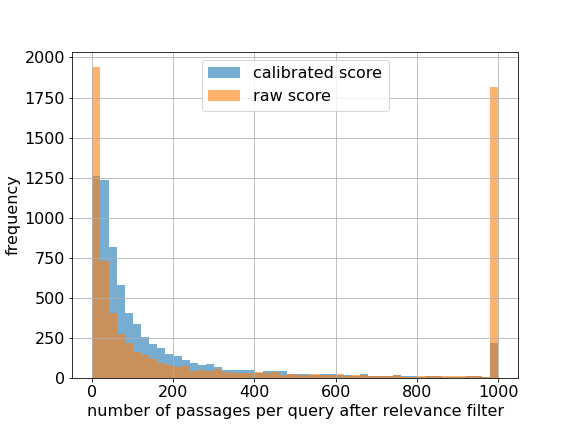}
\caption{Passage count per query after applying filter on calibrated score (power transformation) and raw score. K=1000. Without filtering, passage count is always 1000.}
\label{fig.passage_count_per_query} 
\end{figure}

To illustrate the cross-query comparability of calibrated scores,  we analyze how many passages are retained per query after filtering, as shown in Figure~\ref{fig.passage_count_per_query} ($K=1000$). When filtering on raw scores, there are two prominent spikes in the distribution: at 0 and 1000. This indicates that a global threshold based on raw scores is ineffective, either removing all results or retaining all results for many queries. In contrast, applying a global threshold to the calibrated scores results in a more balanced distribution. This suggests that the calibrated scores enable a more consistent and effective filtering process, ensuring that a reasonable number of relevant results are retained for each query.

Since the filter threshold is tuned such that the recall is 95\%, some relevant items do get filtered out. We have inspected the queries that lose relevant items, and find that they tend to involve rare words (e.g., "what is einstein neem oil good for") and misspellings (e.g., "sydeny climate").

\begin{table*} 
\centering
\small
\begin{tabular}
{p{9cm}|p{0.8cm}|p{0.8cm}|p{0.8cm}|p{1.1cm}|p{0.8cm}|p{1.1cm}} 
 \hline
 \multirow{2}{*}{Passage} & \multirow{2}{*}{Label} & \multirow{2}{*}{Rank} & \multicolumn{2}{c}{K=10} & \multicolumn{2}{|c}{K=1000}\\ 
 \cline{4-7}
  &   &  & Raw score & Calibrated score & Raw score & Calibrated score \\ 
 \hline
  \hline
Average \textbf{Right At Home hourly pay} ranges from approximately $\$$7.36 per hour for Caregiver/Companion to $\$$36.48 pers hour for Registered Nurse - Infusion... & 1 & 1 &  &  & &\\
 \hline
The average salary for work from home jobs is $\$$64,000...  & 0  & 2 &  & $\times$ & &\\ 
 \hline 
According to Salary.com, the average stay-at-home mom or dad should be earning $\$$134,121.00 per year for performing ten common daily functions... & 0 & 3 &  & $\times$ & &\\
 \hline
Right At Home Salaries in United States. Salary estimated from 1,967 employees, users, and past and present job advertisements on Indeed in the past 12 months. & 0  & 4 &  & $\times$ & &\\ 
 \hline
Gain Financial Freedom \& Earn $\$$10.00 - $\$$125.00 per hour work at home! ... & 0 & 5 &  & $\times$ & & \\ 
 \hline
Industry Avg. About The Price Is Right Play at Home Game TV Commercial, 'Win $\$$4000 Cash'... & 0  & 6 & & $\times$ & & \\ 
 \hline
The average KB Home salary ranges from approximately $\$$39,845 per year for Purchasing Coordinator to $\$$118,625 per year for Director... & 0 & 7 & & $\times$ & &\\ 
\hline
The average KB Home salary ranges from approximately $\$$39,845 per year for Purchasing Coordinator to $\$$121,644 per year for Director. .. & 0  & 8 &  & $\times$ & &\\ 
\hline
Average Homegoods hourly pay ranges from approximately $\$$7.50 per hour for Customer Service Representative to $\$$25.00 per hour for Industrial Mechanic... & 0 & 9 &  & $\times$ & & $\times$\\ 
\hline
Work-at-home billers are generally offered much lower pay, around $\$$0.10 to $\$$0.25 per billing claim...  & 0  & 10 &  & $\times$ & & $\times$ \\ 
\hline
 \end{tabular}
\caption{Query is "how much does right at home pay". The calibrated score results are based on power transformation. $\times$ denotes the passage is filtered out.}
\label{table.msmarco_example}
\end{table*}

\subsubsection{Case study}

To further illustrate how our relevance filtering method performs, we showcase an example for the query "how much does right at home pay". As shown in Table~\ref{table.msmarco_example}, only the first passage among the retrieved top-10 is relevant. Filtering based on raw scores struggles to differentiate relevant passages from irrelevant ones, retaining 10 and 366 passages when retrieving $K=10$ and $K=1000$ candidates respectively. In contrast, filtering using our power-transformed score demonstrates a stark improvement. For $K=10$, it retains only the single relevant passage, effectively filtering out all irrelevant ones. Even when scaling up to $K=1000$, only 7 irrelevant passages are kept. This example reinforces our previous findings: by transforming raw scores to a more interpretable scale, our method allows a global threshold to effectively filter out irrelevant candidates.

\subsection{Experiment on Walmart product search data}

\begin{table} 
\centering
\begin{tabular}
{p{2cm}|p{2.8cm}|p{1cm}|p{1cm}}
 \hline
 Dual encoder objective & Method  & PR AUC & P@R95    \\ 
 \hline
  \hline
  - & no filter & - &  P=0.4916, R=1.0  \\
  \hline
 \multirow{5}{*}{Listwise loss} 
 & raw score & 0.7074 & 0.5217 \\ 
 & raw score (max-norm) & 0.7374 & 0.5865 \\ 
 \cline{2-4}
 & linear & 0.8590 & 0.6139 \\ 
 & square root  & 0.8588 & \textbf{0.6231}  \\ 
 & quadratic & 0.8609 & 0.6160 \\ 
 & power  & \textbf{0.8619} & 0.6225  \\ 
 \hline
 \multirow{5}{*}{Contrastive loss}  
  & raw score  & 0.7889 & 0.5702  \\ 
  & raw score (max-norm) & 0.7159 & 0.5711  \\ 
  \cline{2-4}
 & linear & \textbf{0.8626} & \textbf{0.6361}  \\ 
 & square root  & 0.8606 & 0.6302  \\ 
 & quadratic  & 0.8623 & 0.6330  \\ 
 & power & 0.8567 & 0.6247  \\ 
\hline
 \end{tabular}
\caption{Experiments on Walmart product search dataset. The best scores are shown in boldface. "P" and "R" denote for precision and recall, respectively.}
\label{table.wmt_offline}
\end{table}

\begin{table} 
\centering
\begin{tabular}{p{0.5cm}|cp{1cm}} 
 \hline
  K & lift (p-value) \\ 
  \hline
 5 & +5.34\% (0.00) \\
 \hline
 10  & +4.00\% (0.00) \\
 \hline
\end{tabular}
\caption{Precision of the top K results for impacted queries.}
\label{table.online_relevance}
\end{table}

\begin{table} 
\centering
\begin{tabular}{l|l|l} 
 \hline
   & Orders & GMV \\ 
  \hline
 lift (p-value) & +0.03\% (0.86) & -0.11\% (0.83) \\
 \hline
\end{tabular}
\caption{A/B test results}
\label{table.online_engagement}
\end{table}

\label{section.wmt_offline}
We conduct similar experiments with Walmart product search data. 
The training dataset consists of 700k queries and 6 million query-product pairs, each assessed by human reviewers for relevance and categorized as "exact", "substitute", or "irrelevant", similar to the Amazon shopping dataset \cite{reddy2022shopping}.

We compare the performance of filtering methods using the PR AUC and P@R95 metrics defined above. In addition to the differences between various calibration functions, we are also interested in how our methods perform on dual encoders trained with different objectives: contrastive loss and listwise loss. Our previous experiments showed that models trained with listwise loss usually perform better in retrieval tasks \cite{magnani2022semantic}.

\begin{table*} 
\centering
\begin{tabular}
{p{0.7cm}|p{11cm}|p{1.7cm}|p{3cm}} 
 \hline
Rank & Products & Label & Presence with filtering \\ 
 \hline
  \hline
1	& Starbucks Paradise Drink Pineapple Passionfruit with Coconut Milk, 14 fl oz Bottle	& Exact	&  \\
\hline
2	& Starbucks Pink Drink Strawberry Acai Coffee Coconut Milk Drink, 14 fl oz Bottle	& Substitute	& \\
\hline
3	& Starbucks Paradise Drink Pineapple Passionfruit with Coconut Milk, 14 fl oz, 12 Pack Bottles	& Exact &	\\
\hline
4	& Welch's Passion Fruit Fruit Juice Drink, 59 fl oz carton	& Irrelevant	&  \\
\hline
5	& Starbucks Refreshers Peach Passion Fruit Sparkling Juice Blend, 12 Oz Can	& Substitute	&  \\
\hline
6	& Welch's Passion Fruit Fruit Juice Drink, 89 fl oz carton	& Irrelevant	& $\times$ \\
\hline
7	& Welch's Passion Fruit Juice Cocktail, 64 fl oz Bottle	& Irrelevant	&  $\times$ \\
\hline
8	& Goya Pineapple Juice, 33.8 fl oz	& Irrelevant	& $\times$ \\
\hline
9	& Dole Pineapple 100\% Juice 59 oz Bottle	& Irrelevant	& $\times$ \\
\hline
10	& Starbucks Refreshers Sparkling Juice Blend, Peach Passion Fruit With Coconut Water, 12 oz Can	& Substitute & \\
\hline
 \end{tabular}
\caption{Query is "starbucks pineapple passion fruit juice". $\times$ denotes the product is filtered out.}
\label{table.walmart_example}
\end{table*}

\subsubsection{Implementation}
We train two dual-encoder models using contrastive loss (Equation~\ref{eq.contrastive_loss}) and listwise loss (Equation~\ref{eq.softmax_loss}), respectively. The dual encoder training follows \cite{magnani2022semantic}, where DistilBERT \cite{sanh2019distilbert} is utilized for both query and product encoders. During Cosine Adapter training, the dual encoder is frozen. 
The "exact", "substitute", and "irrelevant" query-product pairs are assigned labels of 1, 0.5, and 0, respectively. (We find that setting "substitute" to 0.5 leads to better performance, because "substitute" is more relevant than "irrelevant"). The models were trained with binary cross entropy loss (Equation~\ref{eq.loss}) until convergence utilizing 4 A100 GPUs with a batch size of 512. Note that the dual encoders are trained on engagement data, while the Cosine Adapter is trained on manually annotated relevance data.

\subsubsection{Results}
The test dataset contains 1000 queries with 10 products each. The query-product pairs are evaluated by human reviewers in the same manner as the training data. When computing metrics, we treat "exact" products as positive and "substitute" and "irrelevant" products as negative.

The results in Table \ref{table.wmt_offline} show that our proposed methods outperform the baselines. The improvement is particularly pronounced when applied to scores from the listwise-loss-trained dual encoder. For listwise loss, square root and power transformations perform the best. For constrastive loss, linear transformation performs the best.

\subsubsection{Analysis}
When filtering on raw cosine scores, the contrastive-loss models outperform the listwise-loss models by 9\% in P@R95. This is expected as listwise loss prioritizes relative ranking within a candidate list, potentially leading to less calibrated raw scores, even though they contain important information about relative relevance. However, by applying the Cosine Adapter, contrastive loss's lead is narrowed to 2\%.

Like in Section 5.2.3, since the filter is tuned to have recall=95\% across all queries, some relevant items products end up being filtered out. From inspection of the results of the power-transformation filter, the queries with the lowest recall show a pattern of containing rare brand names, numbers, and/or misspellings.

\subsection{Experiment on Walmart search system}
\subsubsection{Setup}
To demonstrate the practical impact of our proposed relevance filtering component, we integrated it into Walmart's embedding-based retrieval system, following the architecture outlined in Figure~\ref{fig.workflow}. Walmart's search system utilizes a hybrid retrieval approach, combining traditional lexical retrieval and embedding-based retrieval \cite{magnani2022semantic}. We evaluated our proposed method on top of the production baseline, assessing its impact via two measures: relevance, assessed by human judgment of the top 10 products surfaced to customers (post-reranker), and engagement, measured through an online A/B test. The model deployed was the square root Cosine Adapter; the filtering threshold was calibrated for a 99\% recall to minimize the risk of filtering out relevant products. 

\subsubsection{Results}
To evaluate the relevance performance, we sampled about 700 impacted queries (queries with different ranking in the top 10 positions), and assessed the relevance of the top 10 products surfaced to the customers following the same rating guideline as detailed in Section~\ref{section.wmt_offline}. We then computed precision by treating the "exact" products as positive. The results are presented in Table~\ref{table.online_relevance}. 
While the combined testing with downstream product reranking might partially dilute the relevance lift, we still observe a notable improvement in precision: over 5\% for top 5 and 4\% for top 10. The results of our A/B test are presented in Table~\ref{table.online_engagement}, showing the impact on the number of orders and Gross Merchandise Value (GMV). Overall, we observe a neutral impact on engagement, with p-values much greater than 0.05 for both metrics. Thus, our online tests demonstrate an improvement in precision without negatively impacting engagement.


\subsubsection{Case study}
To illustrate the impact of relevance filtering module in Walmart's production retrieval system, we show the top 10 products for the query "starbucks pineapple passion fruit juice" in Table~\ref{table.walmart_example}. Without filtering, only 20\% of the top-10 products are relevant. Among the irrelevant products, 80\% exhibit flavor mismatches, while 50\% show brand mismatches. These errors mainly came from embedding-based retrieval, which doesn't rely on keyword matching. However, by incorporating the relevance filtering module, we successfully filtered out 80\% of the irrelevant products in the top 10.  

\section{Conclusion}
In this paper, we propose a novel relevance filtering component for embedding-based retrieval systems that enhances precision at the expense of a small loss of recall. Our approach involves transforming raw cosine similarity scores into interpretable relevance probabilities using query-dependent mapping functions. This method is computationally efficient and easy to implement, requiring only a lightweight adapter module added to the existing dense retrieval system. Extensive evaluation, including experiments on both the public MS MARCO dataset and a private Walmart product search dataset, along with a live A/B test on Walmart.com, demonstrates the effectiveness of our approach in improving precision across diverse applications.

\section{Resources}
Code, checkpoints and instructions are available at \url{https://github.com/juexinlin/dense_retrieval_relevance_filter}.

\bibliographystyle{ACM-Reference-Format}
\balance
\bibliography{references}

\end{document}